\documentclass[%
 reprint,
superscriptaddress,
nofootinbib,
 amsmath,amssymb,
 aps,floatfix,
]{revtex4-1}
\usepackage{color}
\usepackage{graphicx}
\usepackage{graphics}
\usepackage{caption}
\usepackage{bigints}
\usepackage{multirow}
\usepackage{subcaption}
\usepackage{dcolumn}
\usepackage{bm}
\RequirePackage{mathptmx}      
\RequirePackage{amsmath,amssymb, amsbsy}
\RequirePackage{bm}
\usepackage{adjustbox}
\RequirePackage{bbm}
\usepackage[ruled]{algorithm2e}
\RequirePackage{flushend}
\RequirePackage{natbib}
\usepackage[colorlinks,citecolor=blue,urlcolor=blue,linkcolor=blue]{hyperref}
\RequirePackage[usenames, dvipsnames]{xcolor}
\RequirePackage{tikz}
\RequirePackage[percent]{overpic}
\RequirePackage{setspace}
\usetikzlibrary{shapes.geometric, arrows}
\usepackage{natbib}

\newcommand{\rpm}{\sbox0{$1$}\sbox2{$\scriptstyle\pm$}
  \raise\dimexpr(\ht0-\ht2)/2\relax\box2 }

\usepackage{bbold}
\usepackage{xfrac}

\usepackage{orcidlink}

\usepackage{dcolumn}
\usepackage{bm}

\newcommand{\Real}{\mathbb{R}}

\newcommand{\wn}{w_{_N}\hspace{-0.025cm}}
\newcommand{\vn}{v_{_N}\hspace{-0.025cm}}
\newcommand{\un}{u_{_N}\hspace{-0.025cm}}
\newcommand{\thetahat}{\widehat{\mathbf{\theta}}}
\newcommand{\Chat}{\widehat{\mathbf{C}} }

\newcommand{\bmtheta}{\mathbf{\theta}}

\newcommand{\residuals}{ \widehat{\mathbf{\epsilon}}}

\newcommand{\CM}{\mathbf{C}^M(\bmtheta)}
\newcommand{\CMdot}{\dot{\mathbf{C}}^M(\bmtheta)}
\newcommand{\bmr}{\mathbf{r}}
\newcommand{\bmrtilde}{\widetilde{\mathbf{r}}}

\begin{document}


\title{Testing Models for Angular Power Spectra: A Distribution-Free Approach
}
\thanks{This letter has been submitted in conjunction with its companion manuscript \cite{PRD}}.%
\author{Sara Algeri\,\orcidlink{0000-0001-7366-3866}}
\email{salgeri@umn.edu}
\affiliation{School of Statistics, University of Minnesota, Minneapolis, MN,  USA}
\author{Xiangyu Zhang\,\orcidlink{0000-0003-3684-0370}}
\affiliation{School of Statistics, University of Minnesota, Minneapolis, MN,  USA}
\author{Erik Floden\,\orcidlink{0000-0002-1701-7461}}
\affiliation{School of Physics and Astronomy, University of Minnesota, Minneapolis, MN,  USA}
\author{Hongru Zhao\,\orcidlink{0000-0002-1753-4709}}
\affiliation{School of Statistics, University of Minnesota, Minneapolis, MN,  USA}
\author{Galin L. Jones\,\orcidlink{0000-0002-6869-6855}}
\affiliation{School of Statistics, University of Minnesota, Minneapolis, MN,  USA}
\author{Vuk Mandic\,\orcidlink{0000-0001-6333-8621}}
\affiliation{School of Physics and Astronomy, University of Minnesota, Minneapolis, MN,  USA}
\author{Jesse Miller\,\orcidlink{0009-0005-9465-7461}}
\affiliation{School of Statistics, University of Minnesota, Minneapolis, MN,  USA}

\date{\today}

\begin{abstract}
A novel goodness-of-fit strategy is introduced for testing models of angular power spectra with unknown parameters. Using this strategy, it is possible to assess the validity of such models without specifying the distribution of the angular power spectrum estimators. This holds under general conditions, ensuring the method's applicability in diverse applications. Moreover, the proposed solution overcomes the need for case-by-case simulations when testing different models, leading to notable computational advantages.
\end{abstract}

\maketitle

\section{Motivation and goal}
\label{sec:intro}
The angular power spectrum characterizes the power distribution of quantities on a sphere across angular scales. Understanding its features and assessing the validity of the corresponding theoretical models is critical in studying a variety of fundamental phenomena in different scientific fields. For instance, in geodesy it is used to characterize the properties of the Earth's sea surface topography \cite{hwang} as well as the gravitational and magnetic fields of Earth and other planets \cite{Albertella1999-vr, Simons_2006, Lesur2006-ey, livermore2024reconstructions}. It is also used in geophysics to infer mechanical properties of the lithosphere \cite{wieczorek}, in atmospheric science to examine surface temperature anomalies \cite{Evans, Sneppen_2022}, and in medical imaging applications \cite{Maniar2005TheCP, MitraM06}. 
Of particular interest to us is its relevance in cosmology and astrophysics.
For instance, in the context of the cosmic microwave background (CMB), angular power spectra of both CMB temperature and polarization have enabled precise measurements of essential cosmological parameters, such as the energy density of dark matter and dark energy ~\cite{planck2020}. Galaxy surveys, like the Sloan Digital Sky Survey (SDSS) \cite{SDSS_DR16}, have provided insights into the galaxy distribution across the sky. The angular power spectrum derived from these distributions has shed light on the evolution of visible matter in the universe \cite{Hayes_SDSSAngSpec,yang2023}. Similarly, weak gravitational lensing surveys, such as the Dark Energy Survey (DES), have generated weak lensing convergence sky-maps and their angular power spectrum describes the mass distribution in the universe \cite{DESY3LensingMap,DESY3_DM}. Measurements of the stochastic gravitational-wave (GW) background energy density across the sky and its angular power spectrum could be used to identify the astrophysical or cosmological mechanism(s) that generated it
\cite{Geller:2018mwu,Cusin:2017mjm,Cusin:2019jpv,Cusin:2017fwz,Jenkins:2018uac,Jenkins:2018lvb}, albeit to date only upper limit constraints have been made\textcolor{black}{~\cite{Thrane:2009fp, o1_directional,Abbott:O2_aniso,O3stochdir,O3_ASAF,deepali_targeted, PhysRevLett.122.081102, PhysRevD.100.063527}}. Additionally, cross-correlations of these GW \textcolor{black}{background} sky maps and their electromagnetic counterparts (galaxy counts, weak lensing, etc.) have been proposed as auxiliary astrophysical probes \cite{Cusin:2017mjm,Cusin:2019jpv,Cusin:2017fwz}, and first measurements have been made~\cite{yang2023} \footnote{\textcolor{black}{While there are studies considering cross-correlations between electromagnetic signals and distributions of individually resolvable GW sources} \textcolor{black}{ \cite{Calore_2020, zazzera2024gravitationalwavesgalaxiescrosscorrelations, PhysRevD.111.063513, PhysRevLett.116.121302, PhysRevD.93.083511, PhysRevD.94.023516, Scelfo_2018},  we refer specifically to a GW background signal.}}. This trend is anticipated to persist and intensify with the release of new telescopes and detectors in the upcoming decade, with a tenfold (or greater) surge in the acquired data volumes.

Let $T(\widehat\Omega,\mathbf{x})$ be a sky map with $\widehat\Omega$ being the sky direction and $\mathbf{x}$ denoting other \textcolor{black}{independent variables}, such as redshift or frequency bin of the observed light or GWs. For example, $T(\widehat\Omega,\mathbf{x})$ may represent the number of galaxies observed over different pixels and at a given redshift, or the CMB temperature measured across the sky and at a given CMB frequency bin. \textcolor{black}{Note that $\mathbf{x}$ can be multidimensional, i.e., $\mathbf{x}\in \mathcal{X}\subseteq \Real^D$.} 

Consider the decomposition of $T(\widehat\Omega,\mathbf{x})$ given by
\begin{eqnarray}
\label{eqn:expansion}
    T(\widehat\Omega,\mathbf{x}) & =  \sum_{\ell=0}^{\infty} \sum_{m = -\ell}^{\ell} a_{\ell m} (\mathbf{x}) Y_{\ell m }(\widehat\Omega) 
\end{eqnarray}
with $Y_{\ell m }(\widehat\Omega)$ being the spherical harmonics, that is, complex-valued basis functions defined on a sphere, and 
\begin{eqnarray}
\label{eqn:coeff}
a_{\ell m}(\mathbf{x})=  \int T(\widehat\Omega,\mathbf{x}) Y_{\ell m} (\widehat\Omega ) d\widehat\Omega.
\end{eqnarray}
The angular power spectrum is defined as 
\begin{eqnarray}
\label{eqn:Al}
    C_\ell (\mathbf{x}) & = & \frac{1}{2\ell + 1} \sum_{m = -\ell}^{\ell} |a_{\ell m}(\mathbf{x})|^2.
\end{eqnarray}

In practical applications, the sky map $T(\widehat\Omega,\mathbf{x})$ is often directly measured in the pixel basis and can then be converted into estimates $\widehat{a}_{\ell m}(\mathbf{x})$ using \eqref{eqn:expansion}. In some cases, the available data covers only a limited portion of the sky, hence resulting in large experimental uncertainties for $\widehat{a}_{\ell m}(\mathbf{x})$ estimators at low $\ell$'s. 
In other cases, such as when measuring the anisotropic stochastic GW background, it is possible to directly estimate the $a_{\ell m}(\mathbf{x})$ coefficients (and the corresponding covariance matrix) by cross-correlating the time-series strain data from two geographically separated GW detectors~\cite{allenromano,Thrane:2009fp,O3stochdir}.
In any case, by replacing $a_{\ell m}(\mathbf{x})$ with $\widehat{a}_{\ell m} (\mathbf{x})$ in \eqref{eqn:Al}, and for any value of $\mathbf{x}$ fixed, we can  obtain  estimates $\widehat{C}_{\ell} (\mathbf{x})$, $\ell=1,\dots, L$ with $L$ being the point of truncation of the expansion in \eqref{eqn:expansion}. Hereafter assume the latter to be unbiased estimates of $C_{\ell} (\mathbf{x})$ for all $\ell=1,\dots, L$ (see \citet{PRD} for details on the construction of unbiased estimators).  \textcolor{black}{Note that while $\ell=0$ can be included, we choose not to do so since often one is interested only in fluctuations of $T(\widehat\Omega,\mathbf{x})$ around its mean, implying $C_0(\mathbf{x})=0$. 
}

Let $\mathbf{C}^{M} (\mathbf{x},\bmtheta)$ be a theoretical model with components $C^{M}_\ell (\mathbf{x},\bmtheta)$ aiming to predict $C_\ell (\mathbf{x}) $ in \eqref{eqn:Al}. Denote with $\widehat{\mathbf{C}}(\mathbf{x})$ the vector with components $\widehat{C}_{\ell} (\mathbf{x})$. Once an estimate of the unknown parameter   $\bmtheta\in \Real^p$ is obtained, our goal is to assess the validity of $\mathbf{C}^{M} (\mathbf{x},\bmtheta)$ based on the information carried by $\widehat{\mathbf{C}}(\mathbf{x})$ and its covariance matrix $\bm{\Sigma}(\mathbf{x})$.

Various methods for estimating  $\mathbf{C}^{M} (\mathbf{x},\bmtheta)$ have been proposed and often rely on the assumption that, for each $\mathbf{x}$ fixed, $\widehat{\mathbf{C}}(\mathbf{x})$ is a Gaussian random vector \cite{Netterfield_2002, PhysRevLett.85.1366, Piacentini_2006, Jungman_1996, Knox_1995}. 
However, while the estimates $\widehat{a}_{\ell m}(\mathbf{x})$ are usually obtained by averaging large amounts of data and are asymptotically distributed as a complex multivariate Gaussian, the $\widehat{C}_{\ell}(\mathbf{x})$ are averages of the squares of $\widehat{a}_{\ell m}(\mathbf{x})$. It follows that their distribution is that of a generalized $\chi^2$ whose likelihood function does not have a closed-form expression \cite{PRD} and, in general, cannot be considered approximately Gaussian. 
Solutions that aim to account for non-Gaussianity typically involve likelihoods that are combinations of a Gaussian and offset log-normal distribution \cite{Bennett_2003, Bennett_2013}. However, those have also been shown to be unsatisfactory in some settings \cite{ref3}. 
Unfortunately, when the distribution of $\widehat{\mathbf{C}}(\mathbf{x})$ is not adequately modeled, constructing a goodness-of-fit test for $\mathbf{C}^{M}(\mathbf{x},\bmtheta)$  is particularly challenging. 
Nonetheless, such a test is required to ensure the reliability of the resulting estimates of $\bmtheta$ and subsequent conclusions.  

\textcolor{black}{Motivated by this challenge, we discuss a novel distribution-free approach to testing theoretical models for angular power spectra.  Traditionally, distribution-free goodness-of-fit tests are defined as tests that rely on test statistics with an asymptotic null distribution independent of the model or distribution being tested. 
Here, however, it acquires a broader meaning: the null distribution of the test statistics used does not depend on the model. Additionally, to derive this distribution, one does not need to know the distribution of the random variables whose mean is being tested, in our case, $\widehat{\mathbf{C}}(x)$. This is made possible by reformulating the estimation problem of $\mathbf{C}^{M}(\mathbf{x},\bmtheta)$ as a regression problem and implementing the testing procedure proposed by
\citet{khm21}. This letter is the first work applying such a technique to enable distribution-free testing of models for the angular power spectrum. }

\textcolor{black}{
A significant computational advantage is that a distribution-free testing solution avoids the need for case-by-case simulations when testing different models. It is, therefore, especially desirable when testing models with a complicated structure, as in the case of astrophysical anisotropic stochastic GW backgrounds or when
cross-correlating different sky maps such as galaxy count maps and GW background sky maps~\cite{Cusin:2017mjm, Cusin:2019jpv, Jenkins:2018uac, Cusin:2017fwz}.  }


This letter provides a self-contained exposition of the method in general settings. 
In a companion paper \citep{PRD}, we present all of the technical details for the main results, discuss their extensions, and illustrate the implementation of the methodology to study the anisotropic stochastic gravitational-wave background. Finally, while the proposed procedure is motivated by the need for distribution-free inference for autocorrelation angular power spectra, the statistical tools described here also apply to the analysis of cross-correlation angular power spectra.

\textcolor{black}{\section{Estimation via Generalized Least Squares}}
\label{sec:estimation}
\textcolor{black}{
Let $\mathbf{x}_1, \dots,  \mathbf{x}_n$ be the vectors collecting the values of the independent variables. We assume that such values are linearly ordered so that, if $\mathcal{X}\subseteq \Real$, $\mathbf{x}_1\leq \mathbf{x}_2 \leq \dots\leq \mathbf{x}_n$. When $\mathcal{X}\subseteq \Real^D$, we assume that the $D$-dimensional vectors $\mathbf{x}_i=(x_{i1},\dots,x_{iD})$  are ordered by increasing one dimension at a time. For example, if $D=2$ and $n=n_1\cdot n_2$, then:
\begin{equation}
\begin{split}
\mathbf{x}_1&=(x_{11},x_{12}), \mathbf{x}_{2}=(x_{21},x_{12}),\dots, \mathbf{x}_{n_1}=(x_{n_11},x_{12}),\\
&\mathbf{x}_{n_1+1}=(x_{11},x_{22}),\dots,\mathbf{x}_{2n_1}=(x_{n_11},x_{22}),\text{ etc.}
\end{split}
\end{equation}} 

Denote with  $\widehat{\mathbf{C}} $ the vector of components $\widehat{C}_{\ell}(\mathbf{x}_i)$, with $\ell=1,\dots,L$ and $i=1,\dots,n$,  i.e., if $\widehat{\mathbf{C}}^T(\mathbf{x}_i)=[\widehat{C}_1(\mathbf{x}_i),\ldots,\widehat{C}_L(\mathbf{x}_i)]$, then $\widehat{\mathbf{C}}$ is a vector of dimension  $N=L \cdot n$ and  defined as
\begin{equation}
    \label{Chat}
    \
    \widehat{\mathbf{C}} =[\widehat{\mathbf{C}}^T(\mathbf{x}_1),\ldots,\widehat{\mathbf{C}}^T(\mathbf{x}_n)]^T.
\end{equation}
For example, $\widehat{\mathbf{C}}(\mathbf{x}_i)$ may represent the estimated angular power spectrum up to the $L$-th  multipole for the GW energy density at the  GW frequency $\mathbf{x}_i$. Hence, $\widehat{\mathbf{C}}$ collects the estimated angular power spectra at frequencies  $\mathbf{x}_1,\dots,\mathbf{x}_n$.
The vector $\mathbf{C}^M(\bmtheta)$ of elements $C^M_{\ell}(\mathbf{x}_i,\mathbf{\theta})$ are specified similarly.  

Let $\bm{\Sigma} $ be the covariance matrix of $\widehat{\mathbf{C}} $. We assume the random vectors $\widehat{\mathbf{C}}(\mathbf{x}_i)$ at different $\mathbf{x}_i$ values to be independent so that, for all $\ell,\ell'=1,\dots,L$ and $i\neq i'$,
\begin{equation}\text{cov}\bigl(\widehat{C}_\ell(\mathbf{x}_i),\widehat{C}_{\ell'}(\mathbf{x}_{i'})\bigl)=0.\end{equation}
Therefore, $\bm{\Sigma} $ is a block diagonal matrix with blocks $\bm{\Sigma}(\mathbf{x}_i)$. 
We further assume that $\bm{\Sigma} $ is either known exactly, well approximated, or estimated by means of a consistent estimator \cite{PRD}. As an example, in the context of the anisotropic stochastic GW background, the covariance matrix for the $\widehat{a}_{\ell m}(\mathbf{x}_i)$ estimators can be estimated directly from the time-series strain data for each frequency bin $\mathbf{x}_i$~\cite{allenromano, Thrane:2009fp, O3stochdir} and the estimators in different frequency bins are independent to a very good approximation \textcolor{black}{\citep{Thrane:2009fp,allenromano}. This covariance matrix can then be used to construct the block diagonal matrix $\bm{\Sigma} $, with each block corresponding to a different GW frequency $\mathbf{x}_i$.}

Estimating the astrophysical parameter $\bmtheta$ characterizing $\CM$ can be accomplished using least squares estimation.
That is, by solving
\begin{equation}
\begin{split}
\label{optim}
    \thetahat=\arg\min_{\bmtheta}(\Chat-\mathbf{C}^M(\bmtheta))^T\bm{\Sigma}^{-1} (\Chat-\mathbf{C}^M(\bmtheta)).
\end{split}
\end{equation}
Notice that the optimization problem in \eqref{optim}, known in the statistical literature as `Generalized Nonlinear Least Squares' \citep[e.g.,][Ch. 6-7]{davidson} is equivalent to using maximum likelihood for estimating $\bmtheta$ when the likelihood of $\Chat$ is approximately Gaussian \citep[e.g.,][]{ref1,ref2,ref3}. However, the rationale behind \eqref{optim} is simply that of minimizing the sum of the 
squares of the decorrelated  residuals collected in the vector \textcolor{black}{
\begin{equation}
\label{residuals}
   \residuals=[\widehat{\epsilon}_1,\dots,\widehat{\epsilon}_N]^T=\bm{\Sigma}^{-1/2}(\Chat-\mathbf{C}^M(\thetahat))
\end{equation}}
where $\bm{\Sigma}^{-1/2}$ is the inverse square-root matrix of $\bm{\Sigma}$ obtained, for example, via Jordan or Schur decomposition \citep[e.g.,][]{higham}. 
\textcolor{black}{The estimator $\thetahat$ is consistent based on the classical theory of $M$-estimation \citep{huber}. That is, regardless of the distribution of $\Chat$, as $N\rightarrow\infty$, $\thetahat$ converges to the true value of $\theta$ when $\mathbf{C}^M(\theta)$ is correctly specified. Such a property is also necessary to enable the validity of the testing approach described in \ref{sec:GOF} (Cf. \cite{PRD} for the technical details). }

\textcolor{black}{
It follows that what is relevant to ensure the consistency of $\thetahat$ as an estimator of $\bmtheta$ is not the Gaussianity of  $\Chat$ but rather the validity of the postulated model $\CM$. Therefore, our goal is to test the hypothesis:
\begin{equation}
    \label{test}
    H_0:  \langle\Chat \rangle =\CM.
\end{equation}}

\section{Distribution-free goodness-of-fit}
\label{sec:GOF}
\textcolor{black}{The $N\times 1$ vector of decorrelated residuals, $\residuals$ in \eqref{residuals}, allow us to quantify the discrepancy between the estimated angular power spectrum, $\Chat$, and the proposed model $\CM$. To test the hypothesis in \eqref{test}, we consider the process of partial sums of the decorrelated residuals
\begin{align}
    \label{partial_sums}
    \wn(t)=\frac{1}{\sqrt{N}}\sum_{k=1 }^t\widehat{\epsilon}_{k}=\frac{1}{\sqrt{N}}\residuals^T\mathbb{I}_t 
\end{align}
where  $t=1,\dots,N$ indexes the elements of the process and
$\mathbb{I}_t=[\mathbb{1}_{\{1\leq t\}},\mathbb{1}_{\{2\leq t\}},\dots,\mathbb{1}_{\{N\leq t\}}]^T$.
Hence,
\begin{equation}
w_{_N}(1)=\frac{\widehat{\epsilon}_{1}}{\sqrt{N}},w_{_N}(2)= \frac{\widehat{\epsilon}_{1}+\widehat{\epsilon}_{2}}{\sqrt{N}},\dots,w_{_N}(N)=\frac{1}{\sqrt{N}}\sum_{k=1}^N\widehat{\epsilon}_{k}. 
\end{equation}
} 
 A family of test statistics for \eqref{test} is easily constructed by  
considering functionals of $\wn(t)$. For example,  by choosing
\begin{equation}
\label{functionals}
\begin{split}
h_{\text{KS}}\bigl[\wn(t)\bigl] &= \max_t |\wn(t)|
\end{split}
\end{equation}
we obtain  the Kolmogorov-Smirnov  statistic \citep[]{darling}  adapted to our context. 

\textcolor{black}{While the elements of $\residuals$ are uncorrelated, it is possible that some form of nonlinear dependence may persist among them when $\Chat$ is not approximately Gaussian. In most cases, however, this does not represent an impediment to deriving the limiting distribution of $\wn(t)$. In particular, if the dependence is `local' as is the case, for example, when the vectors
$\widehat{\mathbf{C}}^T(\mathbf{x}_i)$, introduced in Section \ref{sec:estimation}, are independent from one another, under $H_0$ and as $N\rightarrow\infty$, the process $\wn(t)$ can be shown to converge to a  Gaussian process with mean zero and covariance function approximately equal to:
\begin{equation}
\label{eqn:cov_wt}
\frac{1}{N}\mathbb{I}^{T}_{t}[\mathbf{I}_N-\sum_{j=1}^p\mathbf{\mu}_j\mathbf{\mu}^T_j\bigl]\mathbb{I}_{t'}
\end{equation}
with $\mathbf{I}_{_N}$ denoting the $N\times N$ identity matrix and the set of vectors $\{\mathbf{\mu}_j\}_{j=1}^p$ is such that: }
\begin{equation}
\label{matrix}
\left[\begin{array}{c}
\mathbf{\mu}^T_1\\
\vdots\\ 
\mathbf{\mu}^T_p 
\end{array}\right]=\left[\CMdot^T\bm{\Sigma}^{-1} \CMdot\right]^{-1/2}\left[\bm{\Sigma}^{-1/2}\CMdot\right]^{T},
\end{equation}
where $p$ is the size of $\mathbf{\theta}$ and  $\CMdot=\frac{\partial \CM}{\partial \bmtheta}$. \textcolor{black}{The structure of the vectors $\mathbf{\mu}_j$ strictly depends on the type of estimator employed    (see \cite{PRD} for more details). }

\textcolor{black}{
From \eqref{eqn:cov_wt}, it follows that the asymptotic null distribution of $\wn(t)$ depends on the model being tested through its covariance. As a result, one must simulate the null distribution of the desired test statistic on a case-by-case basis for each of the models being tested with Monte Carlo methods or the parametric bootstrap\footnote{\textcolor{black}{The parametric bootstrap is a resampling method that 
differs from classical Monte Carlo methods in that various samples are simulated by replacing the true value of $\bmtheta$ with its estimator $\thetahat$ \cite{efron}.}}. 
Unfortunately, this may be computationally expensive when testing a number of different, and possibly complicated, models. 
Nevertheless, it is possible to overcome this limitation by constructing a new set of residuals and a corresponding process of partial sums, with limiting null distribution independent of  $\CM$.} 

The main building block is the operator
\begin{equation}
    \label{eqn:U}
U_{\mathbf{a},\mathbf{b}}\mathbf{q}=\mathbf{q}-\frac{\langle\mathbf{a}-\mathbf{b},\mathbf{q}\rangle}{1-\langle\mathbf{a},\mathbf{b}\rangle}(\mathbf{a}-\mathbf{b}),
\end{equation}
originally introduced in the context of distribution-free goodness-of-fit for discrete distributions \cite{khm13}. Hereafter, we refer to the map entailed by the operator $U_{\mathbf{a},\mathbf{b}}$ as \emph{Khmaladze-2} (or \emph{K2})  \emph{transformation} \footnote{The nomenclature \emph{Khmaladze-2} (or \emph{K2})  \emph{transformation} is used to distinguish the map entailed by the operator $U_{\mathbf{a},\mathbf{b}}$ from the  ``Khmaladze transformation'' or  Khmaladze-1 (K1) transformation \cite{khm82}.}.

Given two vectors, $\mathbf{a}$ and $\mathbf{b}$, with $\|\mathbf{a}\|=\|\mathbf{b}\|=1$, the unitary operator $U_{\mathbf{a},\mathbf{b}}$ maps $\mathbf{q}=\mathbf{a}$ into  $\mathbf{b}$, $\mathbf{q}=\mathbf{b}$ into $\mathbf{a}$, and leaves vectors $\mathbf{q}$ orthogonal to $\mathbf{a}$ and $\mathbf{b}$ unchanged. 
In our setup, $U_{\mathbf{a},\mathbf{b}}$ is employed to map the model-dependent vectors $\{\mathbf{\mu}_j\}_{j=1}^p$, characterizing the covariance of the process $\wn(t)$, into an arbitrarily chosen, model-independent, orthonormal set of vectors $\{\mathbf{r}_j\}_{j=1}^p$.

By adapting the testing framework of \citet{khm21} to our context, we can construct residuals with limiting null distribution depending solely on the vectors $\{\mathbf{r}_j\}_{j=1}^p$ and which is, consequently, independent of the model  $\CM$ being tested. 
The main steps of this approach are now described. 

\begin{itemize}
\item[]\hspace{-0.5cm}\textbf{Step 1. } Estimate $\bmtheta$ via \eqref{optim} and construct the vectors $\mathbf{\mu}_j$  in \eqref{matrix} with $\bmtheta$ replaced by $\thetahat$.

\item[]\hspace{-0.5cm}\textbf{Step 2. } Given an arbitrarily chosen orthonormal set $\{\bmr_j\}_{j=1}^p$, construct a new set of vectors $\{\bmrtilde_j\}_{j=1}^p$ orthogonal to each $\mathbf{\mu}_{j'}$, with $j'<j$, by setting $\bmrtilde_{1}=  \bmr_{1}$ and constructing subsequent vectors $\bmrtilde_j$ as
\begin{equation} \bmrtilde_j=U_{\mathbf{\mu}_{(j-1)},\bmrtilde_{(j-1)}}\dots U_{\mathbf{\mu}_1 \bmrtilde_1}  \bmr_j\end{equation}
for example, 
\begin{equation}
\begin{split}
\bmrtilde_{2}&=U_{\mathbf{\mu}_1 \bmrtilde_1} \mathbf{r}_{2}\\
\widetilde{\mathbf{r}}_{3}&=U_{\mathbf{\mu}_2 \widetilde{\mathbf{r}}_{2}}U_{\mathbf{\mu}_{1} \bmrtilde_{1}} \mathbf{r}_{3}, \text{ etc.}
\end{split}
\end{equation}
Notice that each operator $U_{\mathbf{\mu}_{j},\bmrtilde_{j}}$ is applied to everything on its right-hand side.

\item[]\hspace{-0.5cm}\textbf{Step 3. } Consider the operator 
\begin{equation}
\label{Up}
\mathbf{U}_p^T\mathbf{q}=U_{\mathbf{\mu}_{1} \bmrtilde_{1}} \dots U_{\mathbf{\mu}_{p},\bmrtilde_{p}}\mathbf{q}
\end{equation}
mapping vectors $\mathbf{\mu}_j$ into vectors $\bmr_j$
and construct the vector of \emph{K2 residuals} \textcolor{black}{ \begin{equation}
\widehat{\mathbf{e}}= \mathbf{U}_p^T \Bigl[\residuals-\sum_{j=1}^p\mathbf{\mu}_j\mathbf{\mu}^{\sf T}_j\residuals\Bigl]
=\mathbf{U}_p^T\residuals -\sum_{j=1}^p\mathbf{r}_j\mathbf{r}^{\sf T}_j\mathbf{U}_p^T\residuals
\end{equation}}

\item[]\hspace{-0.5cm}\textbf{Step 4. } Let the  partial sums of the K2 residuals be
\begin{equation}
    \label{rotated_process_p}
   \vn(t)= \frac{1}{\sqrt{N}}\widehat{\mathbf{e}}^{T} \mathbb{I}_t.
\end{equation}
Similarly to \eqref{functionals}, test statistics for \eqref{test} can be constructed by taking functionals $h[\cdot]$ of $\vn(t)$. 

\item[]\hspace{-0.5cm}\textbf{Step 5. } Compute the value of the desired test statistic, $h[\vn(t)]$, and compare it with the distribution of the same functional taken with respect to a zero-mean Gaussian process, $\un(t)$, with covariance function 
\textcolor{black}{
\begin{equation}
\label{eqn:cov}
\begin{split}
\langle \un(t),\un(t') \rangle= \tfrac{1}{N}\mathbb{I}_t^T\biggl[I-\sum_{j=1}^p \mathbf{r}_j\mathbf{r}_j^T\biggl]\mathbb{I}_{t'}.\\
\end{split}
\end{equation}}
\textcolor{black}{The distribution of $h[\un(t)]$ can be derived directly  by simulating realizations of 
\begin{equation}
\label{eqn:approx}
\un(t)=\frac{1}{\sqrt{N}}\mathbb{I}_t^T[\mathbf{e}-\sum_{j=1}^p\mathbf{r}_j\mathbf{r}^{T}_j\mathbf{e}]
\end{equation}
with $\mathbf{e}$ drawn from a multivariate standard normal. }
\item[]\hspace{-0.5cm}\textbf{Step 6. } Reject the model $\CM$ if the probability $P(h[\un(t)]>h[\vn(t)])$ is smaller than the desired significance level.
\end{itemize}

The construction of the $\bmrtilde_j$ vectors in Step 2 is an intermediate step necessary to map vectors $\mathbf{\mu}_j$'s into $\bmr_j$ using the operator $\mathbf{U}_p^T$ \cite{PRD}.   This enables the construction of the K2 transformed residuals, $\widehat{\mathbf{e}}$, as described in Step 3.  
In contrast to the process $\wn(t)$ in \eqref{partial_sums}, under $H_0$,  the process of partial sums of the K2 residuals, $\vn(t)$, constructed in Step 4, has the same limiting distribution as the Gaussian process $\un(t)$. The covariance function of the latter is given in \eqref{eqn:cov} and depends solely on the $-$ arbitrarily chosen $-$ orthonormal set $\{\bmr_j\}_{j=1}^p$. This is true regardless of the distribution of $\Chat$ \cite{PRD}. 
 It follows that, in the limit,  test statistics constructed as outlined in Step 5 have a limiting null distribution that is independent of both the distribution of $\Chat$ and the model $\CM$ being tested. 
Finally, for sufficiently large $N$, the probability in Step 6 provides an asymptotic approximation of the p-value for testing the hypothesis in \eqref{test}.

\begin{figure*}
\includegraphics[scale=0.38]{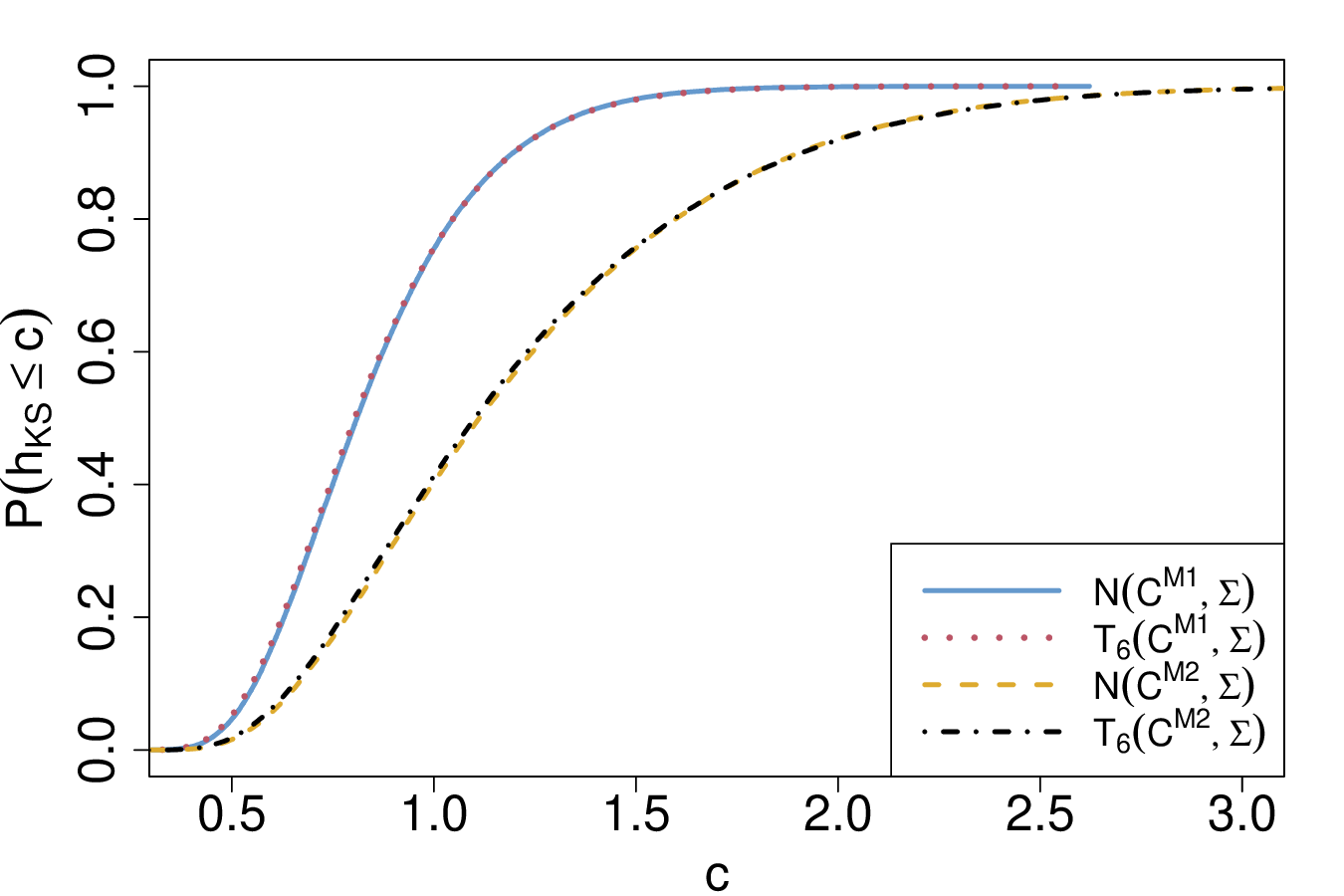}
\includegraphics[scale=0.38]{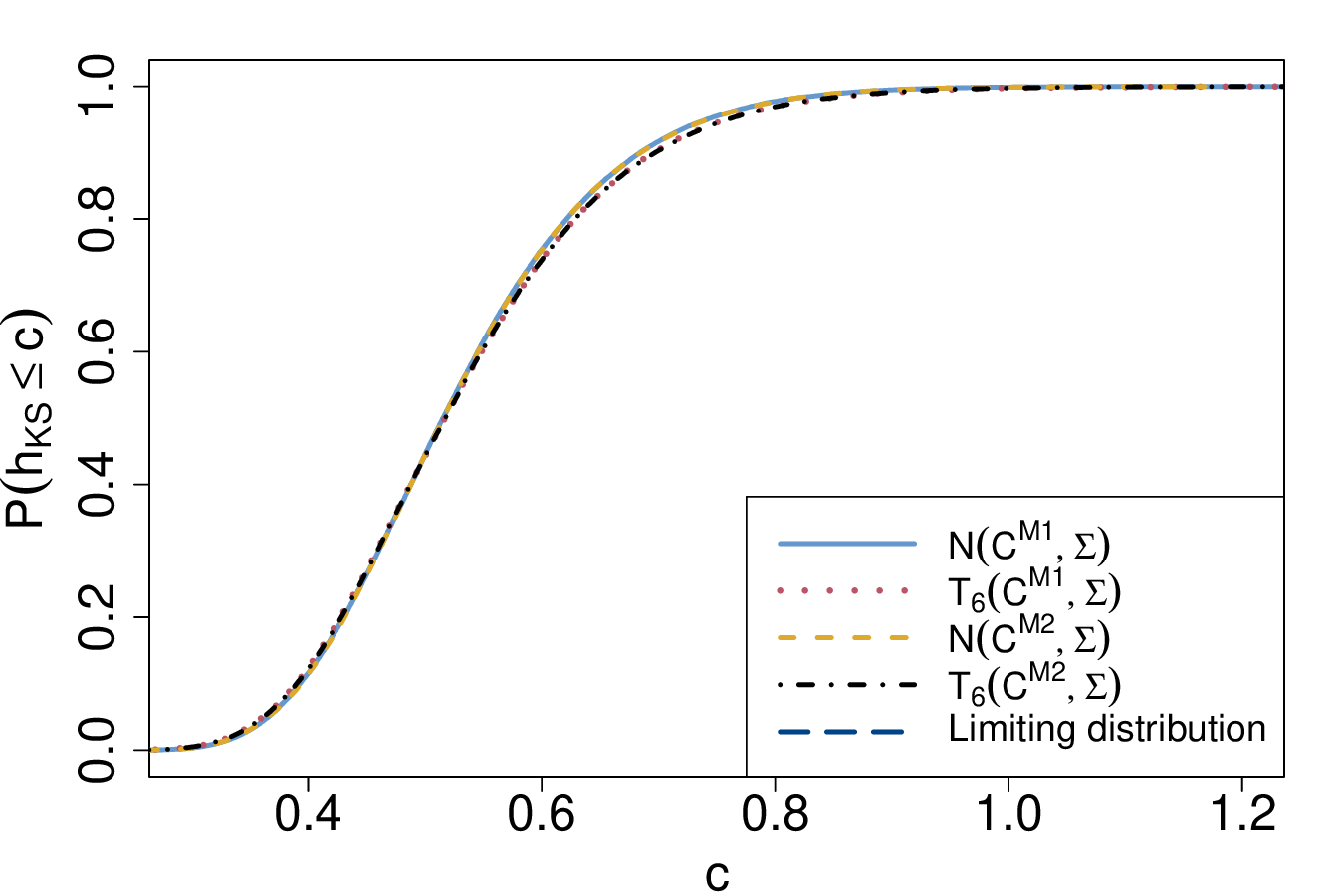}
\caption{Simulated distributions of the Kolmogorov-Smirnov statistic computed as in \eqref{functionals} (left panel)  and \eqref{eqn:KSrotated}  (right panel) for each of the configurations  in \eqref{eqn:models}-\eqref{eqn:distributions}. Each simulation was obtained using 100,000 bootstrap replicates.
}
\label{Fig:1}
\end{figure*}
\textcolor{black}{
\section{An illustrative example}
\label{sec:example}}
Let $\mathbf{C}^{M_1}(\bmtheta)$ and $\mathbf{C}^{M_2}(\bmtheta)$ be two candidate models for $E[\widehat{\mathbf{C}}]$ with components
\begin{equation}
 \label{eqn:models}
\begin{split}
    &C^{M_1}_{\ell}(\mathbf{x}_i,\bmtheta) = \theta_0 + \theta_1 \ell +\theta_2 \mathbf{x}_i,\\
    &C^{M_2}_{\ell}(\mathbf{x}_i,\bmtheta) = \exp\{\theta_0 + \theta_1 \mathbf{x}_i+ \theta_2 \mathbf{x}_i\ell\}.
    \end{split}
\end{equation}
The $\mathbf{x}_i$ values considered correspond to $n=100$ evenly-spaced points on the interval $[0,1]$, and  $\ell=1,\dots,5$. The true value of the parameter $\bmtheta=(\theta_0,\theta_1,\theta_2)$ is $(5,2,4)$. In the numerical experiments conducted, $\bmtheta$ is treated as unknown and is estimated as in \eqref{optim}. For each of the two models in \eqref{eqn:models}, we consider two possible distributions for the corresponding estimator of the power spectrum:
\textcolor{black}{
\begin{equation}
\label{eqn:distributions}
\begin{split}
    &
\widehat{\mathbf{C}}_{m,1}(\mathbf{x}_i) \sim \mathcal{N}\left(\mathbf{C}^{M_m}(\mathbf{x}_i,\bmtheta),\bm{\Sigma}(\mathbf{x}_i)\right), \\
& \widehat{\mathbf{C}}_{m,2}(\mathbf{x}_i) \sim T_6\left(\mathbf{C}^{M_m}(\mathbf{x}_i,\bmtheta),\bm{\Sigma}(\mathbf{x}_i)\right).
    \end{split}
\end{equation}}
That is, for each $\mathbf{x}_i$, \textcolor{black}{$\widehat{\mathbf{C}}_{m,1}(\mathbf{x}_i)$ and $\widehat{\mathbf{C}}_{m,2}(\mathbf{x}_i)$} are independent random vectors following, respectively, a multivariate normal and multivariate $t$-distribution with six degrees of freedom.  The mean vectors, 
$\mathbf{C}^{M_m}(\mathbf{x}_i,\bmtheta)$, have components $C^{M_m}_{\ell}(\mathbf{x}_i,\bmtheta)$  defined as in \eqref{eqn:models} for $m=1,2$. The covariance matrices $\bm{\Sigma}(\mathbf{x}_i)$ are generated from a Wishart distribution with 10 degrees of freedom. 
The left panel of Figure \ref{Fig:1} shows the simulated distribution function of the Kolmogorov-Smirnov statistic in \eqref{functionals} when the data is simulated from the four configurations described above and $H_0$ in \eqref{test} holds -- that is, the model being tested and the one used to simulate the data are the same. Regardless of  \textcolor{black}{$\widehat{\mathbf{C}}_{m,1}(\mathbf{x}_i)$ and $\widehat{\mathbf{C}}_{m,2}(\mathbf{x}_i)$} being $t$ or Gaussian distributed, the null distribution of $h_{\text{KS}}[\wn(t)]$ coincides when testing the same model. This demonstrates that whether or not $\widehat{\mathbf{C}}$ is Gaussian has no impact on the asymptotic null distribution of $h_{KS}[\wn(t)]$. As expected, however, the model under study does affect the asymptotic null distribution. 

\textcolor{black}{
Let us now proceed to implement the procedure described in Steps 1-6. The orthonormal set $\{\mathbf{r}_j\}_{j=1}^p$ selected is such that 
\begin{equation}
\label{r1}
\bmr_1=\biggl[\frac{1}{\sqrt{N}},\dots,\frac{1}{\sqrt{N}}\biggl],
\end{equation}
and subsequent components are constructed via Gram-Schmidt orthonormalization of powers of a vector $\bmr_2$ with elements
\begin{equation}
\label{r2}
r_{2k}=\sqrt{\frac{12}{N}}\biggl[\frac{k}{N}-\frac{N+1}{2N}\biggl].
\end{equation}}

The right panel of Figure \ref{Fig:1} shows the simulated null distribution of the Kolmogorov-Smirnov statistic based on the K2 residuals,
\textcolor{black}{
\begin{equation}
\label{eqn:KSrotated}
h_{\text{KS}}\bigl[\vn(t)\bigl] = \max_t |\vn(t)|,
\end{equation}}
for each of the four scenarios considered. All four curves effectively overlap with the simulated distribution of $h_{\text{KS}}\bigl[\un(t)]$ (blue-dashed lined). The latter represents the limit we expect to achieve, confirming the asymptotically distribution-free nature of the test. 

\textcolor{black}{In summary, these results suggest that one can simulate the null distribution of $h_{\text{KS}}[\wn(t)]$ assuming that $\widehat{\mathbf{C}}$ is Gaussian, even when it is not, without affecting the result of the test. Such a distribution, however, will differ when testing different models, and thus would require a separate simulation for each of them. Alternatively, one could proceed with a statistical test based on $h_{\text{KS}}\bigl[\un(t)]$. In this case, the null distribution would be the same, regardless of the model being tested, and could be easily obtained numerically as summarized in Step 5. } 




\section{Code availability}
\label{sec:code}
The Python code for running the simulations and generating the plots in Section~\ref{sec:GOF}, along with a tutorial for implementing the distribution-free tests in Python, is available at \url{https://github.com/xiangyu2022/DisfreeTestAPS}. The \textsf{R} package \textit{distfreereg} \citep[][]{Miller2024}, available at \url{https://cran.r-project.org/package=distfreereg}, contains a general implementation of the goodness-of-fit test. A tutorial for \textsf{R} is available at \url{https://github.com/small-epsilon/GOF-Testing-for-Angular-Power-Spectrum-Models-in-R}.

\section*{Funding}
SA and XZ are grateful for the financial support provided
by the Office of the Vice President for Research \& Innovation at the University of Minnesota. XZ was partially supported by the University of Minnesota Data Science Initiative with funding made available by the MnDrive initiative. SA, GJ, and VM
were partially supported by NSF grant DMS-2152746. The
work of EF and VM was in part supported by NSF grants
PHY-2110238 and PHY-1806630.  

\nocite{*}
\bibliographystyle{unsrtnat}
\bibliography{apssamp}

\end{document}